\documentclass{ifacconf}

\usepackage{graphicx}      
\usepackage{natbib}        
\usepackage{mathrsfs}
\usepackage{amsmath}
\usepackage{latexsym}
\usepackage{color}
\usepackage{bm}
\usepackage{url}
\usepackage{xcolor}
\usepackage{multirow}

\begin{document}

\begin{frontmatter}

\title{Resilience Analysis of Multi-modal Logistics Service Network Through Robust Optimization with Budget-of-Uncertainty} 

\author[First]{Yaxin Pang},
\author[First]{Shenle Pan},
\author[First]{Eric Ballot}

\address[First]{Mines Paris, PSL University, Centre for management science (CGS), i3 UMR9217 CNRS, 75006 Paris, France (e-mail: yaxin.pang@minesparis.psl.eu; shenle.pan@minesparis.psl.eu; eric.ballot@minesparis.psl.eu) }

\begin{abstract}                
Supply chain resilience analysis aims to identify the critical elements in the supply chain, measure its reliability, and analyze solutions for improving vulnerabilities. While extensive methods like stochastic approaches have been dominant, robust optimization—widely applied in robust planning under uncertainties without specific probability distributions—remains relatively underexplored for this research problem. This paper employs robust optimization with budget-of-uncertainty as a tool to analyze the resilience of multi-modal logistics service networks under time uncertainty. We examine the interactive effects of three critical factors: network size, disruption scale, disruption degree. The computational experiments offer valuable managerial insights for practitioners and researchers.
\end{abstract}


\end{frontmatter}

\section{Introduction}
The significance of supply chain resilience has been unprecedentedly spotlighted due to its increasing and critical importance in mitigating the uncertainties inherent in today's supply chain and logistics services. For example, during the COVID-19 pandemic, shipment delays from China to the U.S. spiked by 33 days \citep{statista}. Another recent example is that car maker Tesla had to shut down their factory near Berlin for 2 weeks due to the delays in the supply chain following the Red Sea attack in 2024 \citep{guardian}. The unreliability of logistics services makes supply chains suffer from violent increases in costs, shipping delays, notable product shortages, and consequently surging product prices. 

Supply chain resilience refers to an organization's capability to endure, adapt to, and recover from disruptions, to fulfill customer demand, maintain target performance levels, and sustain operations in vulnerable environments \citep{hosseini2019review}. Resilience analysis is a key step in the framework, where organizations can measure the supply chain reliability and identify the critical components and vulnerabilities within the supply chains, so that corresponding strategies to enhance the system performance can be designed. Contingent upon the progression of disruption events, resilience analysis can be divided into two phases: disruption phase (robustness and redundancy), and recovery phase (resourcefulness and rapidity) \citep{bruneau2003framework, zhou2019resilience}. The focus of this work lies within enhancing comprehension of system performance and resilience during the disruption phase.

Given the significant impact of shipment delay on supply chain performance, we focus on the research problem of global logistics service network under travel time uncertainty for the distribution of perishable products. More precisely, it refers to the well-known transportation service network design problem, which concerns strategic, tactical or operational planning of services and operations to effectively meet demand while ensuring the profitability of the firm \citep{crainic2000service}. We study multi-modal logistics service networks which integrate several transportation modes, i.e., airway, maritime, road, and railway. Furthermore, this work specifically addresses the products of degradation properties, where the value or quality diminishes with time. This study aims to explore the resilience of logistics service networks in the face of travel time uncertainty resulting from disruptions across various scenarios. It also examines the interactive effects of disruption scale, disruption degree, and network size on network resilience. The insights derived from this investigation are intended to inform future research and practical applications in resilience analysis.

This work aims to make the following contributions to related literature. Firstly, we employ robust optimization with budget-of-uncertainty as a tool to analyze the resilience of multi-modal logistics service networks under time uncertainty. The literature shows that robust optimization is commonly used in the problem of robust planning under uncertainties without specific probability distributions, to find the optimal solutions even for the worst-case scenarios, and with possible consideration of budget of uncertainty \citep{bertsimas2003robust}. However, it has been rarely employed in supply chain resilience analysis. The application here is motivated by two reasons. First, during the unforeseen disruption phase, the impact of the event continues to evolve. It is difficult to gather sufficient historical data to establish the probability distribution of parameters, particularly for travel time, to which perishable products are extremely sensitive. Robust optimization will be more appropriate to this case. Second, we suggest applying the budget of uncertainly to initiate a new approach and perspective to take account of the decision-makers' willingness to accept the risk and the trade-offs between robustness and costs through the budget. The approach can also help identify the most critical and vulnerable components of the networks, which is particularly important for multi-modal services. 

Secondly, we build a Mixed Integer Linear Programming (MILP) model to formulate the problem and minimize the total distribution costs in the worst-case scenarios. Numerical experiments using CPLEX solver are conducted to investigate the impacts of physical network structures and disruption scenarios. Moreover, sensitivity studies have been conducted to examine the three critical factors on resilience performance: network size, disruption scale, and disruption degree. 

Section 2 provides a brief review of the related literature. Section 3 elaborates on the problem and introduces a robust MILP model. Section 4 describes the results and discusses on the findings from computational experiments. Finally, Section 5 provides a summary of this study.

\section{literature Review}
Numerous modeling approaches have been developed to evaluate the performance of supply chain and logistics systems in the face of disruptions. Some reviews proposed to classify the relevant modeling approaches into different categories like mathematical or optimization modeling, topological modeling, simulation modeling, probability theory modeling (e.g. Bayesian network), fuzzy logic modeling, and data-driven modeling \citep{zhou2019resilience}. We refer readers to relevant literature reviews on quantitative analysis of supply chain and logistics system resilience: \cite{hosseini2019review}, and \cite{zhou2019resilience}. 


The literature most relevant to our work can be discussed under the categorization. \cite{khalili2017integrated} applied stochastic programming to an integrated production-distribution planning problem with uncertain production capacity, in order to find an optimal solution on the structure of the chain, risk mitigation, and recovery plan. In the second stage of the model, the worst-case scenario is considered to enhance the robustness of the solution. \cite{dui2021resilience} proposed a deterministic mathematical programming model to decide the recovery sequence of the nodes and edges when several or all nodes/edges fail. \cite{yin2023integrated} evaluated the functional and topological resilience of urban transportation network when exposed to different disruptions through traffic flow simulation model. \cite{wu2023resilience} designed a new resilience indicator for transportation network particularly considering disruption of earthquakes and used Monte Carlo simulation to assess the system performance and investigate the feasibility of a proactive mitigation strategy. \cite{bai2023data} identified key nodes using simulation method and studied the impacts of traffic flow on the resilience of global liner shipping network through a disintegration model. Furthermore, the authors proposed an innovative simulation model with tailored rules to evaluate the dynamic resilience of the system.

Robust optimization holds significant promise as a tool for studying supply chain and logistics resilience due to its computational tractability and non-reliance on probability distribution. This is particularly advantageous in scenarios involving unprecedented disruptions, where obtaining probability distributions is challenging due to the lack of historical data and the evolving of the disruption event. However, as noted by \cite{hosseini2019review}, its potential remains underexplored. In the literature, robust optimization has been mostly applied in robust planning at the operational level, such as freight transportation planning. New techniques have been developed to overcome over-conservatism. For example, \cite{zhang2023full} utilized min-max regret robust optimization to devise defensive strategies for an integrated electricity-gas-transportation system during the pre-disaster phase, highlighting its capacity to mitigate decision-maker conservativeness. \cite{pang2023robust} developed a robust optimization model with heuristic algorithm to address logistics service selection problems under time uncertainty. The authors applied budget of robustness proposed by \cite{bertsimas2003robust} to control the conservativeness of the solution. Distinguished from previous research, this paper focuses on network resilience analysis rather than operational planning, which is rarely investigated from the perspective of robust optimization. To bridge this gap, we aim to explore the potential of robust optimization with budget-of-uncertainty to assess network resilience, and investigate the resilience performance with various network structures and disruption profiles.

\section{Problem description and formulation}

\subsection{Problem description}
We consider a multi-modal logistics service network with different transport modes, i.e., railway, waterway, and airway modes in this paper. Each service $\mathrm{s} \in \mathscr{S}$ corresponds to a type of transport mode, a designated route consisting of a sequential of city nodes, transportation costs, and average travel time, which includes average waiting time on each arc where it runs. Notice that, instead of considering the network as: $\mathscr{G = \{N, A\}}$ like in most previous studies, we apply a multi-modal service network $\mathscr{G = \{N, A, V\}}$ to capture the service features in this problem. Set $\mathscr{N}$ is the set of all city nodes $\mathrm{i} \in \mathscr{N}$ in the network, while $\mathscr{A}$ is the set of all arcs $\mathrm{(i, j)} \in \mathscr{A}$ which link two nodes by at least one service. Especially, we introduce the notation of service-arc $\mathrm{v}$ here as an arc corresponding to a specific service with a form of $\mathrm{(s, i, j)}$. The set of all service-arcs is indicated by $\mathscr{V}$.

To simplify, this work considers only one type of perishable product from one shipper. In order to analyze the design of the network, we consider each client is associated with only one order and one destination address. The demand quantity of each order for the product is the average demand of the corresponding client. All the products are released and prepared at the beginning of the scheduling. Prepared orders can either wait in the departure point (warehouse or factory) for transferring or start to transfer immediately depending on the schedule. Outbound date of each order is the first decision to make. Another decision is to find the optimal transport service selection (and combination) to ship each order from the departure point to its corresponding destination before the shelf-life of the product. The orders can be transshipped at any node in the network from any service entering this node to any service departing from this node. Without loss of generality, we make the following basic assumptions:
\begin{itemize}
    \item Degradation rate is constant for the product and the quality or value degrades with time linearly. Besides, we only consider quality loss but not quantity loss.
    \item The order of each client is not allowed to be split during the whole transportation course.
    \item All orders must be delivered to their respective clients before the specified shelf-life; otherwise, the delivery is deemed a failure.
    \item Due date of delivery is set for each order. No on-time delivery will generate penalty cost, with higher rate for late arrival than early arrival.
\end{itemize}

Travel time along each service-arc is an uncertain parameter due to disruptions. We assume that the probability distribution of travel time is unknown, while the deviation range—the maximum and minimum travel time—is known in this study. We use robust optimization to find optimal solutions under even the worst-case scenario. The latter is defined by the latest arrival date of order (that could be after due date but must be prior to the shelf-life date). Robust optimization with budget-of-uncertainty is applied to mitigate the over-conservatism, where decision-makers can employ a budget to limit the number of service-arcs which will be delayed in the worst-case scenario. Overall, the goal of this problem is to schedule, over a given time horizon, the best outbound date (from departure point) and transport service selection for each order, to minimize total distribution costs, which include transportation cost by using the services, transshipment cost and degradation cost along the delivery, and penalty cost for early or late delivery to the destinations.

\subsection{Mathematical formulation}
\subsubsection{Sets and indices}
\begin{itemize}
    \item $\mathscr{N}$: set of city nodes, $i \in \mathscr{N}$
    \item $ \mathscr{A}$: set of arcs from node $\it{i}$ to node $\it{j}$, $(i,j) \in \mathscr{A}$
    \item $ \mathscr{V}$: set of service-arcs, $(\mathrm{s, i, j})\in \mathscr{V}$
    \item $\mathscr{S}$: set of transport services, $\mathrm{s}\in \mathscr{S}$
    \item $\mathscr{C}$: set of clients, $\mathrm{c}\in \mathscr{C}$
    \item $\mathrm{o}$: node of departure point $o\in \mathscr{N}$
    \item $\mathrm{d^{c}}$: destination of client $\mathrm{c}$, $\mathrm{d^{c}}\in \mathscr{N}$
\end{itemize}
\subsubsection{Parameters and data}
\begin{itemize}
    \item $\mathrm{q^{c}}$: demand quantity for the product of client $\it{c}$
    \item $\mathrm{t^{c}_{(s,i,j)}}$: uncertain parameter indicating the travel time (in day) of service-arc $(\mathrm{s, i, j})$ for client $\mathrm{c}$
    \item $\mathrm{\bar{t}_{(s,i,j)}}$: nominal value of travel time (in day) on service-arc $(\mathrm{s, i, j})$
    \item $\mathrm{\hat{t}_{(s,i,j)}}$: maximum deviation (delay) of travel time on service-arc $(\mathrm{s, i, j})$
    \item $\mathrm{f_{(s,i,j)}}$: 1, if service $\it{s}$ runs on arc $(\mathrm{i,j})$; 0, otherwise
    \item $l$: shelf life of the product
    \item $e^{c}$: due date of client $\mathrm{c}$ 
    \item $\mathrm{\varphi_{(s,i,j)}}$: transportation cost of per unit product by service-arc $(\mathrm{s, i, j})$
    \item $\mathrm{\phi^{s_{1}, s_{2}}_{i}}$: transshipment cost of per product from service $\mathrm{s_{1}}$ to service $\mathrm{s_{2}}$ at node $\it{i}$
    \item $\mathrm{\psi}$: degradation cost of per product per day
    \item $\Gamma$: budget of travel time uncertainty
\end{itemize}
\subsubsection{Decision variables}
\begin{itemize}
    \item $\mathrm{w^{c}}$: continuous variable; the outbound time (in day) of the order by client $\it{c}$ 
    \item $\mathrm{x^{c}_{(s,i,j)}}$: binary variable; 1, if client $\it{c}$ uses service-arc $(\mathrm{s, i, j})$; 0, otherwise
    \item $\mathrm{u^{c}_{(s,i,j)}}$: deviation degree of travel time on service-arc $(\mathrm{s, i, j})$ in the worst case for client $\it{c}$
    \item $\mathrm{z^{c, s_{1}, s_{2}}_{i}}$: binary variable; 1, if the order of client $\it{c}$ is sorted for service transshipment at node $\it{i}$ from service $\mathrm{s_{1}}$ to service $\mathrm{s_{2}}$; 0, otherwise
    \item $\mathrm{k^{c}_{-}}$, $\mathrm{k^{c}_{+}}$: intermediate variables; early and late delivery time of client $\mathrm{c}$, respectively
\end{itemize}

The proposed MILP model is outlined as follows:
\begin{align}
 \min &\hspace{2mm} \Pi = \sum_{(s,i,j)\in \mathscr{A}}\sum_{c\in C}\varphi^{s}_{(i,j)}*q^{c}*x^{c}_{(s,i,j)} + \nonumber\\ 
    &\sum_{i\in \mathscr{N}}\sum_{c\in \mathscr{C}}\sum_{s_{1}\in \mathscr{S}}\sum_{s_{2}\in \mathscr{S}} \phi^{s_{1}, s_{2}}_{i}*q^{c}*z^{c, s_{1}, s_{2}}_{i} + \nonumber\\
    &\psi*\sum_{c\in \mathscr{C}}q^{c}*(w^{c}+\max_{t\in \bm{T}}\sum_{(s,i,j)\in \mathscr{V}}t_{(s,i,j)}*x^{c}_{(s,i,j)}) +\nonumber\\
    & \sum_{c\in \mathscr{C}}(\omega_{e}*q^{c}*k^{-}_{c}+\omega_{a}*q^{c}*k^{+}_{c})
\end{align}

subject to:
\begin{align}
    &\sum_{s\in \mathscr{S}}\sum_{i \in \mathscr{N}}x^{c}_{(s,i,j)}=\sum_{s\in \mathscr{S}}\sum_{i \in \mathscr{N}}x^{c}_{(s,j,i)}, \nonumber\\
    & \hspace{10mm}{\forall c\in \mathscr{C}, j \in \mathscr{N}\setminus \{o, d^{c}\}} \\
    &\sum_{s\in \mathscr{S}}\sum_{i \in \mathscr{N}}x^{c}_{(s,o,i)}=1, {\forall c\in \mathscr{C}}\\
    &\sum_{s\in \mathscr{S}}\sum_{i \in \mathscr{N}}x^{c}_{(s,i,d_{c})}=1, {\forall c\in \mathscr{C}}\\
    &x^{c}_{(s,i,j)} \leq f_{(s,i,j)}, {\forall c\in \mathscr{C}, (s,i,j)\in \mathscr{V}}\\
    &z_{i}^{c,s_{1}, s_{2}} \geq  \sum_{j \in \mathscr{N}}x^{c}_{(s_{1},j,i)} + \sum_{j \in \mathscr{N}}x^{c}_{(s_{2},i,j)} -1,\nonumber \\
    &\hspace{1mm}{\forall c\in \mathscr{C}, s_{1}\in \mathscr{S}, s_{2}\in \mathscr{S}, (s_{1},j,i)\in \mathscr{V}, (s_{2},i,j)\in \mathscr{V}}\\
    &w^{c}+\max_{t\in \mathscr{T}}\sum_{(s,i,j)\in \mathscr{V}}t^{c}_{(s,i,j)}*x^{c}_{(s,i,j)} \leq l, {\forall c\in C}\\
    &w^{c}+\max_{t\in \mathscr{T}}\sum_{(s,i,j)\in \mathscr{V}}t^{c}_{(s,i,j)}*x^{c}_{(s,i,j)}-k_{+}^{c}\leq e^{c}, {\forall c\in C}\\
    &w^{c}+\max_{t\in \mathscr{T}}\sum_{(s,i,j)\in \mathscr{V}}t^{c}_{(si,j)}*x^{c}_{(s,i,j)}+k_{-}^{c}\geq e^{c},{\forall c\in C}\\
    &x^{c}_{(s,i,j)},  z^{c, s_{1}, s_{2}}_{i}\in \{0,1\}, \nonumber\\
    &\hspace{10mm}{\forall c\in  \mathscr{C}, s_{1}\in \mathscr{S}, s_{2}\in \mathscr{S}, (s,i,j)\in  \mathscr{V}, i\in  \mathscr{N}}\\
    &w^{c}, u^{c}_{(s,i,j)}, k^{c}_{-}, k^{c}_{+} \geq 0, {\forall c\in \mathscr{C}, (s,i,j)\in \mathscr{V}}
\end{align}

Objective function (1) aims to minimize total distribution costs including the transportation cost, transshipment cost, cost of degraded product quality, and cost of early and late delivery. Constraints (2)-(4) denote the flow balance constraints. Constraint (5) ensures that the service-arc is selected from available service-arcs. Constraint (6) calculates the auxiliary variable related to transshipment. Constraint (7) guarantees the arrival time of the order for each client is before the shelf life of the product. Constraints (8) and (9) calculate the early and late span of delay respectively. Constraints (10) and (11) define the domain of the variables.

Given nominal value and deviation range of travel time on service-arcs, we formulate the time uncertainty set, which is used in the service time related constraints, i.e., constraints (1), and (7) - (9), as follows by applying robust optimization with budget-of-uncertainty \citep{bertsimas2003robust}: 
\begin{align}
     \nonumber &\mathscr{T} = \{\bm{t}:  t^{c}_{(s,i,j)} = \bar{t}_{(s,i,j)} + u^{c}_{(s,i,j)}*\hat{t}_{(s,i,j)}, \\ 
    &\nonumber \hspace{10mm}\sum_{(s,i,j)\in \mathscr{V}}u^{c}_{(s,i,j)} \le \Gamma,\\
    &\hspace{10mm} 0\leq u^{c}_{(s,i,j)}\leq 1, {\forall (s,i,j) \in \mathscr{V}, c\in \mathscr{C}} \}
\end{align}

The non-linear worst-case scenario part:
\begin{align}
     \max_{t\in \mathscr{T}}\sum_{(s,i,j)\in \mathscr{V}}t^{c}_{(s,i,j)}*x^{c}_{(s,i,j)}
\end{align}
in constraints (1), (7), and (8) can be linearized by the method in \cite{bertsimas2003robust} as formula (14) along with constraints (15) - (16).
\begin{align}
    & \min \Gamma*\lambda^{c}+\sum_{(s,i,j)\in \mathscr{V}}(\bar{t}_{(s,i,j)}*x^{c}_{(s,i,j)}+\theta^{c}_{(s,i,j)})\\
    &\lambda^{c}+\theta^{c}_{(s,i,j)}\geq \hat{t}_{(s,i,j)}*x^{c}_{(s,i,j)},{\forall c\in \mathscr{C}, (s,i,j)\in \mathscr{V}}\\
    &\lambda^{c}, \theta^{c}_{(s,i,j)}\geq 0, {\forall c\in \mathscr{C}, (s,i,j)\in \mathscr{V}}
\end{align}

Additionally, constraint (9) is linearized into constraints (17) - (21) by introducing auxiliary variables.
\begin{align}
    &w^{c}+\sum_{(s,i,j)\in \mathscr{V}}(\bar{t}_{(s,i,j)}*x^{c}_{(s,i,j)}+\hat{t}_{(s,i,j)}*ux^{c}_{(s,i,j)})\nonumber\\
    &\hspace{10mm}+k_{-}^{c}\geq e^{c},{\forall c\in \mathscr{C}}\\
     &ux^{c}_{(s,i,j)} \leq x^{c}_{(s,i,j)}, {\forall c\in \mathscr{C}, (s,i,j)\in \mathscr{V}}\\
    &ux_{(s,i,j)}^{c} \leq u^{c}_{(s,i,j)}, {\forall c\in \mathscr{C}, (s,i,j)\in \mathscr{V}}\\
    &ux^{c}_{(s,i,j)} \geq u^{c}_{(s,i,j)}+x^{c}_{(s,i,j)}-1,{\forall c\in \mathscr{C},(s,i,j)\in \mathscr{V}}\\
    &ux^{c}_{(s,i,j)} \geq 0, {\forall c\in \mathscr{C}, (s,i,j)\in \mathscr{V}}
\end{align}

\section{Numerical experiments}

\subsection{Case setting}
 To investigate the impacts of size of the service network, i.e., the number of service-arcs $|\mathscr{V}|$ in the network, we designed three service networks, indicated as $\mathscr{G}_{1}$, $\mathscr{G}_{2}$, and $\mathscr{G}_{3}$ respectively, which utilize different transportation modes including railway, maritime way, and airway. We selected 27 and 48 arcs from the China-Europe transportation network to be the basic structure of the three networks. This is to say, $\mathscr{G}_{1}$, $\mathscr{G}_{2}$, and $\mathscr{G}_{3}$ share a common network basic, i.e., the same nodes set and arcs set. The services running on these arcs distinguish among these three service networks. $\mathscr{G}_{1}$ consists of 16 services and 50 service-arcs and most of the arcs are equipped with one service. Based on $\mathscr{G}_{1}$, we added more services to each arc in $\mathscr{G}_{2}$ and $\mathscr{G}_{3}$ so that $\mathscr{G}_{2}$ has 25 services and 100 service-arcs while 38 services and 150 service-arcs for $\mathscr{G}_{3}$. 
 
Five clients are designed by randomly choosing cities from the network as destinations, with the demand quantity determined by a uniform distribution $U[5,10]$. Experiments are implemented with 1, 3, and 5 clients respectively, selected sequentially from the designed clients set. The shelf life of the product is set as 30 days. The transportation cost and time associated with each service-arc are considered related to the service mode and distance of the arc. The distance ($km$) between each two nodes ranges between $[300, 12500]$. Given the distance, transportation time is calculated according to the speed of the service mode: 800-1000 $km/h$ for airway, 50-70 $km/h$ for railway, and 46.3-50.0 $km/h$ for waterway, while transportation cost per $km$ is 1-2 euros, 0.15-0.30, and 0.05-0.20, respectively. The transshipment cost at each node respects uniform distribution $U[10,15]$ if the two services are the same transport modes; $U[15,25]$ otherwise. Degradation cost of one unit of product per day is 10\% of the product value, which is determined as 100 here. Early and late delivery cost per unit of product per client per day are 15\% and 20\% of the product value respectively. 

\subsection{Results with sensitivity analysis}
We conducted computational experiments on both small and large-scale cases to validate the effectiveness of the MILP model proposed in this paper and investigate the resilience performance of the transportation networks. The model was programmed in MATLAB R2022a and executed by IBM ILOG CPLEX 12.10.0.0 on a DELL computer equipped with an Intel Core i7-10700 CPU processor (2.90 GHz) and 64.0 GB of RAM. 

The experiment comprises 225 instances, varying in network size ($|\mathscr{V}|$), the proportion of uncertain service-arcs ($PUV$) in the network, deviation rate, and client size. More precisely, $|\mathscr{V}|\in\{50, 100, 150\}$, $PUV$ and deviation rate take value from $[0\%, 25\%, 50\%, 75\%, 100\%]$. To keep the neutral impact of the decision-maker's optimism, $\Gamma$ is set as $50\%*|\mathscr{V}|$. For example, in Instance2 (50\_25\_25), the network comprises 50 service-arcs, where 25\% of the network ($PUV = 25\%$), i.e., 13 services-arcs, have travel time uncertainty, and the deviation rate is 25\% which means the maximum delay $\hat{t}$ of each uncertain service-arc is $25\%*\bar{t}$. All the instances can be solved to optimal in 5 hours. The sensitivity experiments on 1-client, 3-clients, and 5-clients cases are shown in Table~\ref{tb:1client}, Table~\ref{tb:3clients}, and Table~\ref{tb:5clients} respectively. Column $In.$ indicates the instance number, while columns 2 and 3 present the value of $|\mathscr{V}|$ and $PUV$ respectively. The optimal objective function value found for each instance with a particular deviation rate is shown in columns 4-7. The last column indicates the average solving time for the instances with different deviation rates. Note that for the instances with $PUV=0\%$, such as In. 1, 6, 11, etc., the deviation rate has no effect on the optimal results (with no disruption occurred).


\begin{table}[hb]
    \centering
    \caption{Results of 1-client experiments}\label{tb:1client}
    \begin{tabular}{rrr|rrrr|r}
    \hline
        \multirow{2}*{$In.$} & \multirow{2}*{$|\mathscr{V}|$} & \multirow{2}*{$PUV$} & \multicolumn{4}{c|}{deviation rate} & time \\ \cline{4-7}
        ~ & ~ & ~ & 25\% & 50\% & 75\% & 100\% & $100s$ \\ \hline
        1 & \multirow{5}*{50} & 0\%  & 12246  & 12246  & 12246  & 12246  & 0.5   \\ 
        2 & ~ & 25\%  & 12853  & 13461  & 24615  & 25155  & 2   \\ 
        3 & ~ & 50\%  & 12853  & 13461  & 25056  & 25866  & 2   \\ 
        4 & ~ & 75\%  & 13191  & 24426  & 25056  & 25866  & 2   \\ 
        5 & ~ & 100\%  & 14025  & 24786  & 25866  & 153613  & 2   \\ \hline
        6 & \multirow{5}*{100} & 0\%  & 19506  & 19506  & 19506  & 19506  & 1   \\ 
        7 & ~ & 25\%  & 19506  & 19506  & 19506  & 19506  & 5   \\ 
        8 & ~ & 50\%  & 19678  & 20555  & 21433  & 26453  & 5   \\ 
        9 & ~ & 75\%  & 19880  & 20960  & 26392  & 26662  & 5   \\ 
        10 & ~ & 100\%  & 20150  & 21500  & 28588  & 35216  & 5   \\ \hline
        11 &\multirow{5}*{150} & 0\%  & 14847  & 14847  & 14847  & 14847  & 3   \\ 
        12 & ~ & 25\%  & 14847  & 14847  & 14847  & 14847  & 8   \\ 
        13 & ~ & 50\%  & 15426  & 15597  & 15597  & 15597  & 9   \\ 
        14 & ~ & 75\%  & 15619  & 16362  & 17239  & 22443  & 8   \\ 
        15 & ~ & 100\% & 15889  & 17172  & 22376  & 25633  & 12  \\ \hline
    \end{tabular}
\end{table}

\begin{table}[hb]
    \centering
    \caption{Results of 3-clients experiments}\label{tb:3clients}
    \begin{tabular}{rrr|rrrr|r}
    \hline
        \multirow{2}*{$In.$} & \multirow{2}*{$|\mathscr{V}|$} & \multirow{2}*{$PUV$} & \multicolumn{4}{|c|}{deviation rate} &time \\ \cline{4-7}
        ~ & ~ & ~ & 25\% & 50\% & 75\% & 100\% & $100s$ \\ \hline
         16 & \multirow{5}*{50} & 0\%  & 27765  & 27765  & 27765  & 27765  & 2   \\ 
        17 & ~ & 25\%  & 28373  & 28980  & 40135  & 40675  & 12  \\ 
        18 & ~ & 50\%  & 28373  & 28980  & 40576  & 41386  & 12   \\ 
        19 & ~ & 75\%  & 28710  & 39946  & 40856  & 42086  & 12   \\ 
        20 & ~ & 100\%  & 29598  & 41321  & 155852  & 283599  & 49   \\ \hline
        21 & \multirow{5}*{100} & 0\%  & 44299  & 44299  & 44299  & 44299  & 6   \\ 
        22 & ~ & 25\%  & 44299  & 44299  & 44299  & 44375  & 25   \\ 
        23 & ~ & 50\%  & 44470  & 45347  & 46371  & 52021  & 26   \\ 
        24 & ~ & 75\%  & 44672  & 45752  & 51803  & 52777  & 25   \\ 
        25 & ~ & 100\%  & 44942  & 46567  & 54552  & 62192  & 30 \\ \hline
        26 & \multirow{5}*{150}& 0\%  & 31633  & 31633  & 31633  & 31633  & 19   \\ 
        27 & ~ & 25\%  & 31633  & 31633  & 31633  & 31633  & 62   \\ 
        28 & ~ & 50\%  & 32343  & 32993  & 33126  & 33126  & 65   \\ 
        29 & ~ & 75\%  & 32871  & 34561  & 36461  & 43524  & 64   \\ 
        30 & ~ & 100\% & 33952  & 37811  & 45232  & 50354  & 70  \\ \hline
    \end{tabular}
\end{table}
\begin{table}[hb]
    \centering
    \caption{Results of 5-client experiments}\label{tb:5clients}
    \begin{tabular}{rrr|rrrr|r}
    \hline
        \multirow{2}*{$In.$} & \multirow{2}*{$|\mathscr{V}|$} & \multirow{2}*{$PUV$} & \multicolumn{4}{|c|}{deviation rate} &time \\ \cline{4-7}
        ~ & ~ & ~ & 25\% & 50\% & 75\% & 100\% & $100s$ \\ \hline
         31 &  \multirow{5}*{50} & 0\%  & 47233  & 47233  & 47233  & 47233  & 6   \\  
        32 & ~ & 25\%  & 47953  & 48897  & 65484  & 67022  & 25   \\  
        33 & ~ & 50\%  & 52996  & 53654  & 65924  & 70281  & 25   \\  
        34 & ~ & 75\%  & 53334  & 65373  & 67482  & 90053  & 26   \\  
        35 & ~ & 100\%  & 54716  & 79262  & 194293  & 397358 & 29   \\ \hline
        36 &  \multirow{5}*{100} & 0\%  & 74146  & 74146  & 74146  & 74146  & 15   \\  
        37 & ~ & 25\%  & 74146  & 74146  & 74146  & 74146  & 60   \\  
        38 & ~ & 50\%  & 74317  & 75195  & 76848  & 94563  & 61   \\  
        39 & ~ & 75\%  & 74520  & 75825  & 82852  & 95319  & 62   \\  
        40 & ~ & 100\%  & 74790  & 77305  & 98893  & 107272  & 71   \\ \hline
        41 &  \multirow{5}*{150} & 0\%  & 63247  & 63247  & 63247  & 63247  & 62   \\  
        42 & ~ & 25\%  & 64724  & 65581  & 66084  & 66766  & 146   \\  
        43 & ~ & 50\%  & 65011  & 65962  & 66980  & 69760  & 150   \\  
        44 & ~ & 75\%  & 65453  & 67342  & 70084  & 71951  & 146   \\  
        45 & ~ & 100\% & 66045  & 75089  & 79768  & 97523  & 150  \\ \hline
    \end{tabular}
\end{table}

\begin{figure*}
    \begin{center}
\includegraphics[width=18.2cm]{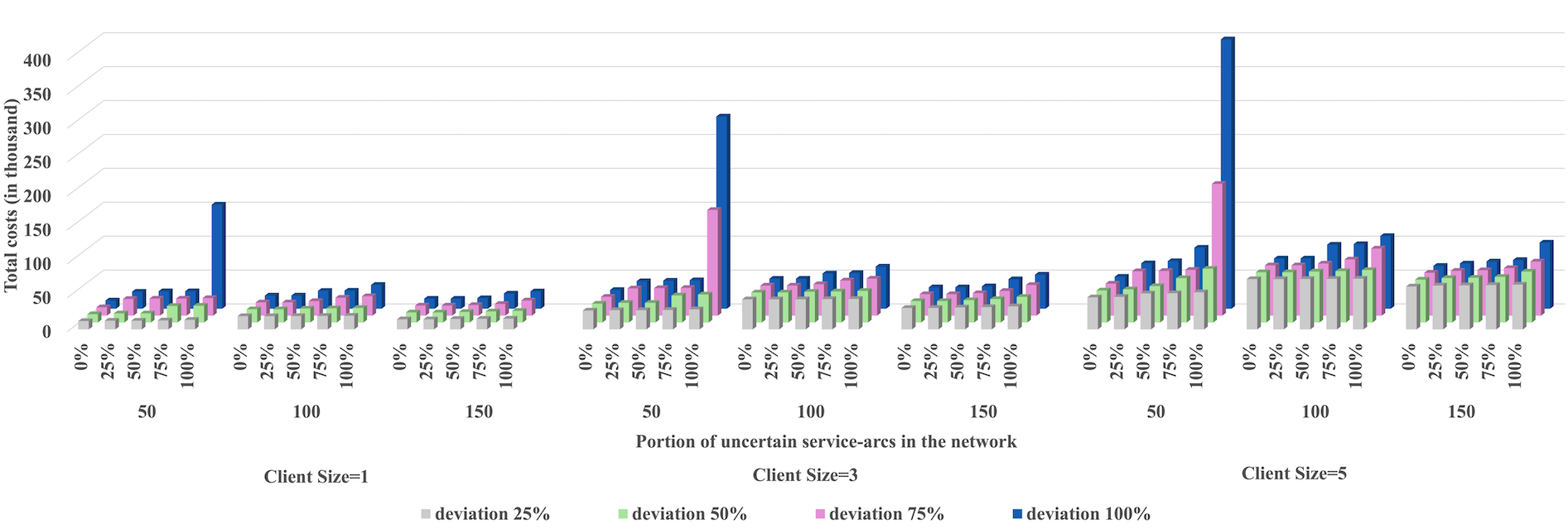}    
\caption{Sensitivity analysis results on network size (50, 100, 150), deviation rate, and proportion of uncertain service-arc} 
\label{fig:sa}
\end{center}
\end{figure*}



Figure ~\ref{fig:sa} illustrates the overall results of Tables 1-3. The first three groups are the results obtained on the instances with only one client, while the three groups in the middle are for 3-clients instances and the last three for 5-clients instances. The resilience performance is evaluated by the total costs change across different instances. Common trends in the change of total costs are explored and analyzed as follows:

\begin{enumerate}
\item Upon examining the results on the instance with a certain client size, it becomes evident that the change of total costs within network $\mathscr{G}_{1}$ are significantly higher compared to network $\mathscr{G}_{2}$ and $\mathscr{G}_{3}$ generally. The reason is $\mathscr{G}_{1}$ has only 50 services-arcs, where the most cases are only one service is available on each arc. This is to say, when disruptions happen and the current solution is not the best any more, the shipper has to choose another substitute arc instead of just changing the service, which might cause higher costs change. $\mathscr{G}_{2}$ and $\mathscr{G}_{3}$ outperform since bigger network size guarantees more alternative service-arc options. Especially, under the disruption profile of 100\% $PUV$ and 100\% deviation rate, the costs by $\mathscr{G}_{1}$ are super high, where the fastest and extremely expensive airway mode has to be used to meet the requests. This further proves the vulnerability of small networks and necessity of improving the network size to an appropriate setting.
\item Within any network among $\mathscr{G}_{1}$, $\mathscr{G}_{2}$ and $\mathscr{G}_{3}$, the total costs exhibit a non-linear growth with an increase on the $PUV$ at a specific deviation rate. The variation of total costs on each interval of $PUV$ is heavily dependant on the service-arcs newly introduced to the uncertainty pool. For example, the instance of 5 clients, within $\mathscr{G}_{2}$ under deviation rate of 100\%, the change of total costs on the $PUV$ interval $[25\%, 50\%]$ is much higher than on the $PUV$ interval $[50\%, 75\%]$. This phenomenon arises because the service-arcs added to uncertainty set of 50\% $PUV$ hold greater significance for the shipper and clients compared to those introduced to the uncertainty set of 75\% $PUV$. This underscores the importance of identifying critical components, i.e., service-arc in this work, in transportation networks. Additionally, this observation can facilitate the identification of critical service-arcs in $\mathscr{G}_{2}$ based on the service-arcs newly introduced to the set of 50\% $PUV$.
\item Within network $\mathscr{G}_{1}$ with one client, the total costs remain manageable under disruption profile of 100\% $PUV$ and 75\% deviation rate, but increase dramatically when deviation rate rises from 75\% to 100\%. This means the robustness degree of network $\mathscr{G}_{1}$ against disruption is 75\%. However, for both 3-clients and 5-clients instances with 100\% $PUV$, the heavy change in total costs occurs when the deviation rate increases from 50\% to 75\%. This indicates the clients set in the 3-clients and 5-clients instances, especially 3-clients, are more sensitive to disruptions. Similarly, given 100\% deviation rate within $\mathscr{G}_{1}$, the costs change of 5-clients instances increases much more when $PUV$ varies from 50\% to 75\% than the ones of 3-clients and 1-client instances. This implies that the 2 clients in the 5-client set are more sensitive to the disruption on the service-arcs introduced to the uncertainty set of 75\% $PUV$ within $\mathscr{G}_{1}$. These findings furthermore confirm the criticality of considering demand distribution in transportation network resilience.
\end{enumerate}

\section{conclusion}
In this work, we explore the potential of robust optimization on resilience analysis and propose a robust optimization model with budget-of-uncertainty for the multi-modal freight logistics service network with travel time uncertainty. Shippers can leverage the proposed model for multiple purposes: first to measure the reliability and resilience of their logistics service network, then to select and plan robust services. This can be achieved by utilizing existing knowledge of the service network structure, service quality information (service time, cost), product information (deteriorating rate, expiration date, if applicable), and the decision-maker's evaluation of the impacts of disruptions. Computational experiments are implemented to investigate the impacts of network structure, disruption scale, and disruption degree on the performance of the service networks. Sensitivity analysis on three critical factors indicates that the network size and demand distribution significantly affect service network resilience. Enhancing resilience by augmenting network size through the identification of critical service-arcs can be highly effective. Moreover, incorporating demand distribution into the analysis can provide more accurate assessment of resilience and identification of critical service-arcs. Nevertheless, further improvements are expected in numerical experiments and sensitivity analysis on the effectiveness of budget, particularly through the integration of real-life cases. Additionally, future research could focus on designing and evaluating various strategies to enhance resilience using the proposed robust optimization methods. Lastly, the proposed model can be compared with stochastic models to further obtain academic insights.

\bibliography{ifacconf}  
\end{document}